\documentclass[pra,twocolumn,superscriptaddress]{revtex4-1}

\usepackage{amsmath}
\usepackage{amssymb}
\usepackage{graphicx}
\usepackage{upgreek}
\usepackage[colorlinks=true, citecolor=blue, urlcolor=blue]{hyperref}

\begin{document}

\title{Experimental realization of Wheeler's delayed-choice experiment with dual selections}
\author{Xiaowan Yang}
\affiliation{School of Physical Science and Technology, Ningbo University, Ningbo, 315211, China}
\author{Xinglei Yu}
\affiliation{School of Physical Science and Technology, Ningbo University, Ningbo, 315211, China}
\author{Liangsheng Li}
\affiliation{National Key Laboratory of Scattering and Radiation, Beijing 100854, China}
\author{Xinzhi Zhao}
\affiliation{School of Physical Science and Technology, Ningbo University, Ningbo, 315211, China}
\author{Tianle Zheng}
\affiliation{School of Physical Science and Technology, Ningbo University, Ningbo, 315211, China}
\author{Chengjie Zhang}
\email{chengjie.zhang@gmail.com}
\affiliation{School of Physical Science and Technology, Ningbo University, Ningbo, 315211, China}
\affiliation{Hefei National Laboratory, University of Science and Technology of China, Hefei 230088, China}
\author{Chuan-Feng Li}
\email{cfli@ustc.edu.cn}
\affiliation{CAS Key Laboratory of Quantum Information, University of Science and Technology of China, Hefei 230026, China}
\affiliation{CAS Center for Excellence in Quantum Information and Quantum Physics University of Science and Technology of China, Hefei 230026, China}
\affiliation{Hefei National Laboratory, University of Science and Technology of China, Hefei 230088, China}
\author{Guang-Can Guo}
\affiliation{CAS Key Laboratory of Quantum Information, University of Science and Technology of China, Hefei 230026, China}
\affiliation{CAS Center for Excellence in Quantum Information and Quantum Physics University of Science and Technology of China, Hefei 230026, China}
\affiliation{Hefei National Laboratory, University of Science and Technology of China, Hefei 230088, China}

\begin{abstract}
{Wheeler's delayed-choice experiment demonstrates wave-particle duality of particles using different experimental configurations
of a Mach-Zehnder interferometer. In a quantum version of this experiment, the wave-particle behavior of photons can be observed by controlling presence or absence of the second beam splitter. Here, we implement a delayed-choice experiment with
dual selections based on entangled photons, experimentally controlling both of two beam splitters in the Mach-Zehnder interferometer, where the presence and absence of the second beam splitter are simultaneously controlled through path encoding. Our
experiment reveals the wave-particle behavior of single photons under different configurations of two beam splitters. We further
discuss the scenario when two beam splitters are in a quantum superposition of being present and absent, photons will be in a
wave-particle quantum superposition state.}
\end{abstract}

\maketitle
\section{Introduction} The principle of complementarity, proposed by Bohr, is one of the central concepts of quantum mechanics \cite{N.Bohr281,Bhuiyan117}. According to this principle, certain properties of a particle are mutually exclusive, such as position and momentum. Particles can exhibit
both particle
and wave characteristics. This is the wave-particle duality, which is a fundamental concept of quantum mechanics~\cite{Yoon21,Li23,Wang21,Chen22,Xue23,Huang23,wen23}. Single-photon Young's double-slit
interference demonstrates the wave nature of photons~\cite{Rueckner131,10}. Some single-photon interference experiments
have also shown that the interference depends on whether the particle goes through both slits or not
~\cite{Aichele052,Grangier863,Jelezko034,Jacques055,Zeilinger056}. On the other hand, the photoelectric effect by Einstein and Compton scattering experiments confirmed the particle nature of light~\cite{Pais798,Compton299}.

In Wheeler's famous delayed-choice experiment~\cite{Jacques0710,BU,BU2,BU3,2,Chaves18,Wheeler1412}, the way we detect photons affects their wave or particle nature, and its outcome depends on whether we know which slit the photon has passed through before it reaches a screen~\cite{Michielsen1013,9,20,Chen21,Xin15,Polino19}. Since then, many related experimental schemes have been reported, such as using single photons and random generators to control their paths, and implementing delayed-choice scenarios in entangled systems~\cite{Jacques0815,Banaszek1316,Ashby0617,4,5,28,Roy12,Vedovato17,Lee14}.
These experiments demonstrate that different detection methods reveal distinct properties of photons.

\begin{figure}
\centering
\includegraphics[width=0.46\textwidth]{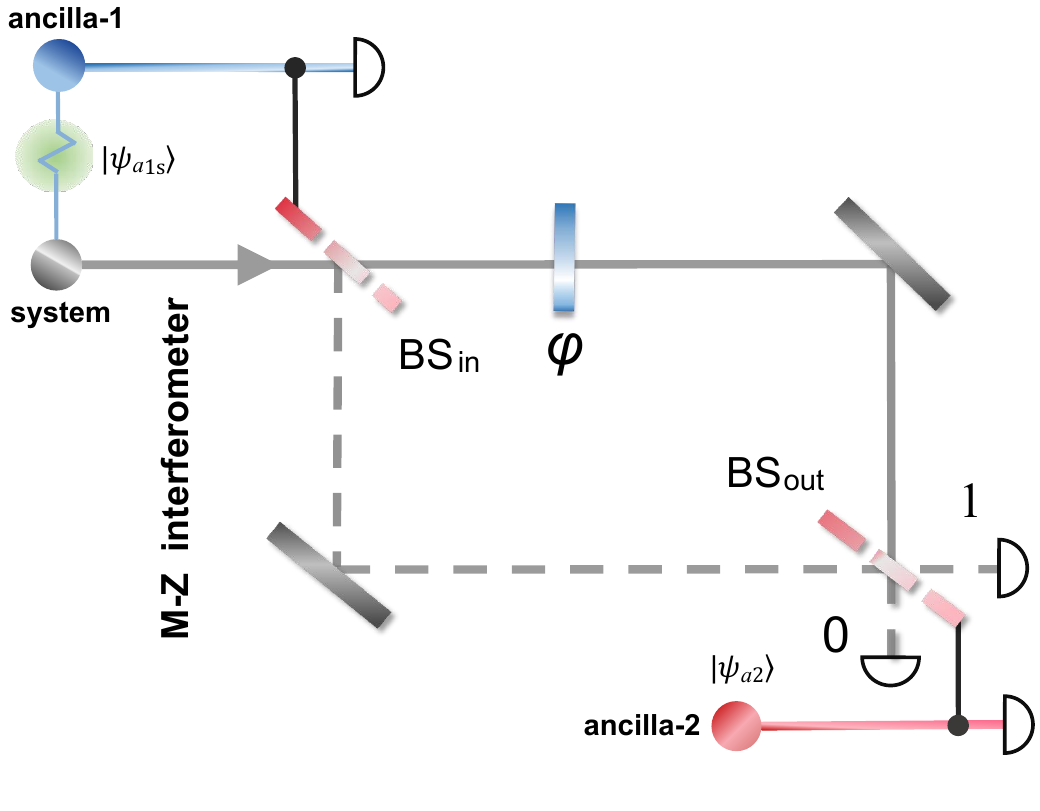}
\caption{A schematic diagram of a dual-selection experimental setup with a Mach-Zehnder interferometer. A tunable phase shifter $\varphi$ is added to one arm of the interference loop, and both of the two beam splitters $\mathrm{BS_{in}}$ and $\mathrm{BS_{out}}$ can be selectively present or absent. The selection of $\mathrm{BS_{in}}$ is controlled by the delayed-choice of measurement basis for ancilla qubit-1 of entangled state $|\psi_{a1s}\rangle=(|0\rangle_{a1}|0\rangle_{s}+|1\rangle_{a1}|1\rangle_{s})/\sqrt{2}$. The state of ancilla qubit-2 is $|\psi_{a2}\rangle=(|0\rangle_{a2}+|1\rangle_{a2})/\sqrt{2}$, and the selection of $\mathrm{BS_{out}}$ depends on the measurement outcomes of ancillary qubit-2.}\label{figure 1}
\end{figure}

Mach-Zehnder interferometer (MZI) is a renowned and efficacious apparatus used to examine wave-particle complementarity, as shown in the system qubit in Figure~\ref{figure 1}. The presence or absence of the output beam splitter ($\mathrm{BS_{out}}$) determines whether the single photon passes through a ``double slit". When $\mathrm{BS_{out}}$ is absent, the detector's response depends on the photon's path and is independent of the relative phase $\varphi$. In this case, only one of the two detectors responds, indicating particle-like behavior.
When $\mathrm{BS_{out}}$ is present, the response probability becomes a  function of $\varphi$, and the photon exhibits wave-like behavior.
To realize the selection of $\mathrm{BS_{out}}$, a postselection scheme utilizing an ancilla qubit was proposed in Ref.~\cite{Ionicioiu1117}. As shown in Figure~\ref{figure 1}, when the ancilla qubit-2 is measured in the basis $\{|0\rangle_{a2}, |1\rangle_{a2}\}$ and yields the outcome $|0\rangle_{a2}$, it indicates the absence of $\mathrm{BS_{out}}$, resulting in the manifestation of particle behavior. Conversely, if the outcome is $|1\rangle_{a2}$, it signifies the presence of $\mathrm{BS_{out}}$, leading to 
wave-like behavior in the particles. Recent studies have extensively explored delayed-choice experiments involving the selection of $\mathrm{BS_{out}}$~\cite{Ionicioiu1117,1,3,G18,W19}, enhancing our understanding of wave-particle duality. Some teams have also used two-photon and three-photon entanglement to perform delayed-choice experiments~\cite{Wang1918,Man1718,Wang2219}. However, there is still a lack of results for delayed-choice experiment with dual selections of two beam splitters (both selection of $\mathrm{BS_{in}}$ and $\mathrm{BS_{out}}$ in Figure ~\ref{figure 1}).

In this work, we investigate  Wheeler's delayed-choice experiment with dual selections using entangled photons. Unlike traditional delayed-choice experiments, which can only select the presence or absence of $\mathrm{BS_{out}}$, dual selections of $\mathrm{BS_{in}}$ and $\mathrm{BS_{out}}$ are implemented,  as shown in Figure~\ref{figure 1}. The selection of $\mathrm{BS_{in}}$ is controlled by the delayed-choice of measurement basis for ancilla qubit-1 of entangled state $|\psi_{a1s}\rangle$, while the selection of $\mathrm{BS_{out}}$ depends on the measurement outcomes of ancilla qubit-2. Our experimental results substantiate the wave-particle duality of single photons, establishing that their wave-like properties emerge solely when a pair of beam splitters is integrated into the MZI setup.

\section{Quantum circuit model}
Quantum circuits provide a clear understanding of the delayed-choice experiment. As depicted in Figure~\ref{figure 2}, the first ancilla qubit and system qubit are in a state of maximal entanglement $|\psi_{a1s}\rangle=\frac{1}{\sqrt{2}}(|0\rangle_{a1}|0\rangle_{s}+|1\rangle_{a1}|1\rangle_{s})$ after the CNOT gate~\cite{Barenco9518,Schwaller2119,Nielsen0120}. When the first ancilla qubit 
is measurement in the basis $\{|0\rangle_{a1}$,$|1\rangle_{a1}\}$ or $\{|+\rangle_{a1},|-\rangle_{a1}\}$, which correspond to the absence or presence of $\mathrm{BS_{in}}$. And the state of ancilla-2 qubit becomes $|\psi_{a2}\rangle=\frac{1}{\sqrt{2}}(|0\rangle_{a2}+|1\rangle_{a2})$ after $H_{2}$ gate, so the application of $H_{3}$ gate is contingent upon the measurement outcome of the second ancilla qubit $-$ either $|0\rangle$ or $|1\rangle$ $-$ indicates the absence or presence of $\mathrm{BS_{out}}$. We denote $P^{(i)}_{jk}$ as the probability that when ancilla qubit-1 is measured in the $\{|0\rangle_{a1},|1\rangle_{a1}\}$ or $\{|+\rangle_{a1},|-\rangle_{a1}\}$ basis with the outcome of $|i\rangle_{a1}$,  the system qubit and ancilla qubit-2 are measured in the $\{|0\rangle_{a2},|1\rangle_{a2}\}$ basis with outcome states being $|j\rangle_{a2}$ and $|k\rangle_{a2}$, respectively.

\section{Experimental setup} The experimental setup of delayed-choice scheme with dual selections is shown in Figure~\ref{figure 3}. It is primarily composed of three parts.

(i) Entangled photon-pair source. A periodically poled ${\rm {KTiOPO_{4}}}$ (PPKTP) crystal, with dimensions of $1 \text{mm} \times 2 \text{mm} \times 20 \text{mm}$, is pumped by a 405 nm continuous-wave laser diode (LD). This setup generates entangled photon pairs via type-II quasi-phase-matched spontaneous parametric down-conversion (SPDC). The resulting photons, known as the signal and idler photons, are entangled~\cite{Cai2224,Kim0325,Shi0426,Lv24}.
Note that the polarization direction of photons will change slightly after passing through single-mode fibers. To correct the polarization of signal photons and idler photons, we introduce two sets of wave plate groups, $\{\mathrm{Q_{1}}, \mathrm{H_{1}}\}$ and $\{\mathrm{Q_{2}},\mathrm{H_{2}}\}$. In our experiment, the horizontally polarized state $|H\rangle$ corresponds to the computational basis state $|0\rangle$, while the vertically polarized state $|V\rangle$ corresponds to state $|1\rangle$. The computational basis states of the idler and signal photons are then encoded as ancilla-1 qubit $\{|0\rangle_{a1},|1\rangle_{a1}\}$ and system qubit $\{|0\rangle_s,|1\rangle_s\}$, respectively.
The half-wave plates $\mathrm{H_{10}}$ and $\mathrm{H_{11}}$ are set at rotation angles of $45^\circ$ and $0^\circ$, respectively. This configuration yields photon pairs in a maximally entangled state $|\psi_{a1s}\rangle=\frac{1}{\sqrt{2}}(|0\rangle_{a1}|0\rangle_{s}+|1\rangle_{a1}|1\rangle_{s}$). 
This is precisely the entangled state used for the selection of $\mathrm{BS_{in}}$, as previously discussed.

\begin{figure}
\centering
\includegraphics[width=0.48\textwidth]{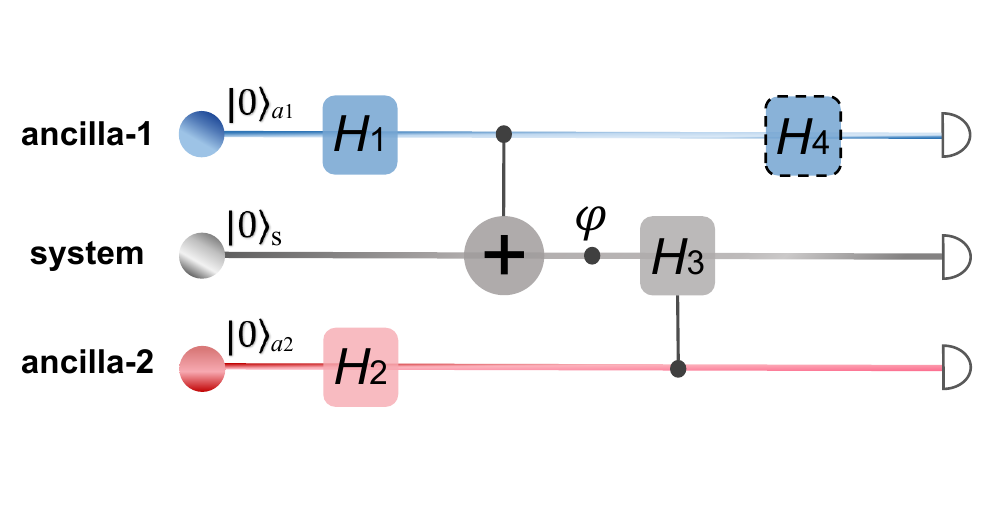}
\caption{By adding two ancilla qubits (represented by blue and red lines), we realize the dual-selection of $\mathrm{BS_{in}}$ and $\mathrm{BS_{out}}$. The first ancilla qubit (blue line) and the system qubit are prepared in a maximally entangled state after passing through the CNOT gate, described by the state $|\psi_{a1s}\rangle$.
$H_{4}$ gate can be optionally placed, and whether $H_{4}$ gate is applied determines the measurement basis of the first ancilla qubit, which realizes the selection of $\mathrm{BS_{in}}$. And the second ancilla qubit controls $H_{3}$ gate, serving as the second controlled-Hadamard gate in Ref. \cite{Ionicioiu1117}. The quantum state of ancilla qubit after passing through $H_{2}$ gate is $|\psi_{a2}\rangle$.
A tunable phase shifter $\varphi$ adds the relative phase $e^{i\varphi}$ to the state $|1\rangle$. $H_{1}$-$H_4$: Hadamard gates.}\label{figure 2}
\end{figure}

\begin{figure*}
\centering
\includegraphics[width=0.8\textwidth]{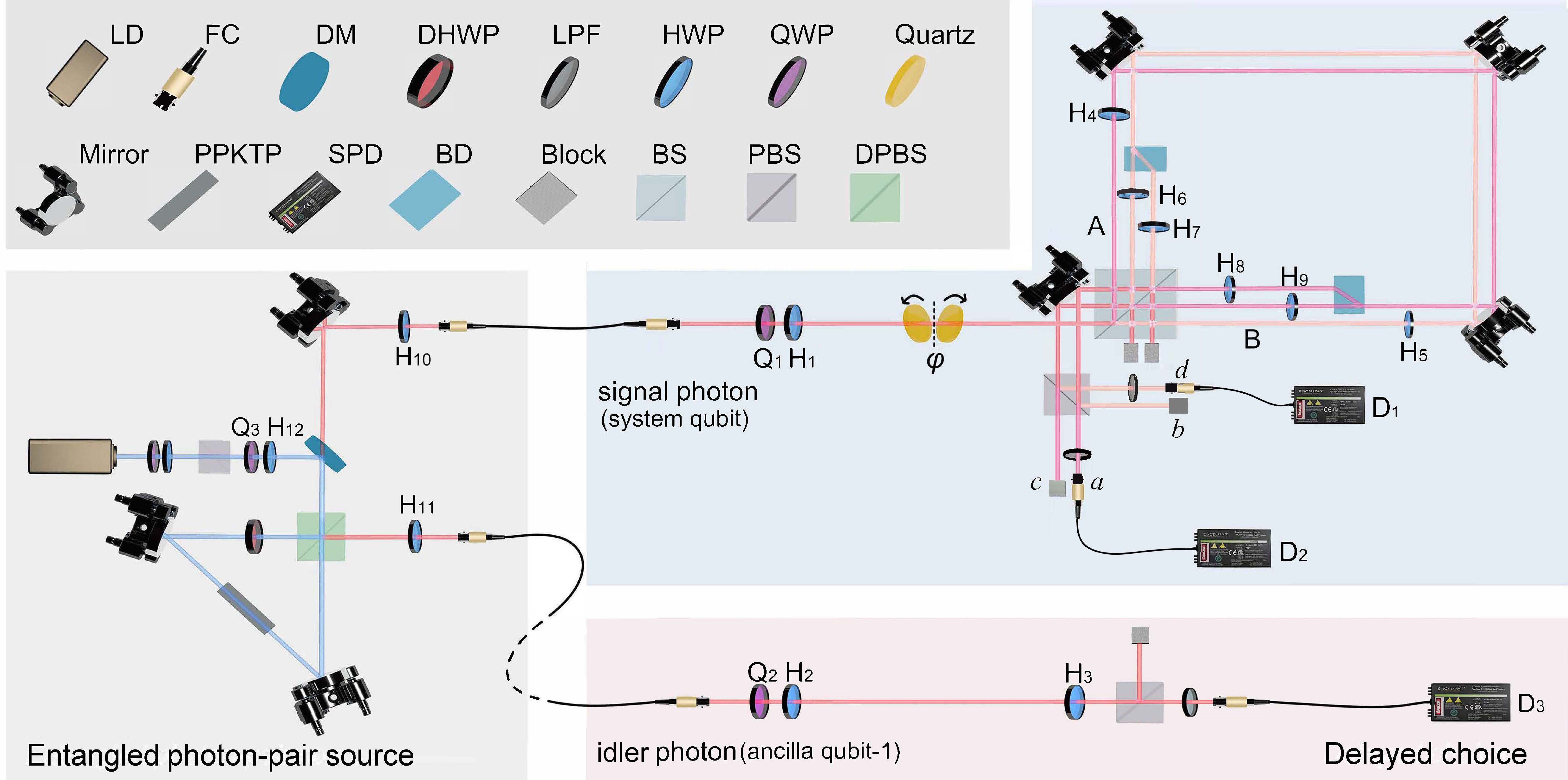}
\caption{Schematic illustration of the experimental setup. Our experimental setup can be divided into three parts: the entangled photon-pair source, delayed-choice and dual selection. In the entangled photon-pair source part, the angle of $\mathrm{H_{12}}$ is $\mathrm{22.5^\circ}$, a photon with a wavelength of 405nm incident on a type-II PPKTP crystal produces a pair of photons with a wavelength of 810nm that are separately transported into single-mode fibers. In the delayed-choice part, $\mathrm{H_{3}}$ and polarizing beam splitter (PBS) in the dotted box comprise the measurement module, $\mathrm{H_{3}}$ has four rotation angles of $\mathrm{0^\circ}$, $\mathrm{45^\circ}$, $\mathrm{22.5^\circ}$ and $\mathrm{-22.5^\circ}$. In the dual-selection part, signal photon passes through quartz crystals which add phase $\varphi$ in vertical polarization state, and then goes into the Sagnac interferometer. The 50:50 non-polarised beam splitter (BS) divides single photon into paths A and B. $\mathrm{H_{4}}$ $(22.5^\circ)$ and $\mathrm{H_{5}}$ $(0^\circ)$ are placed in paths A and B, and polarisation information is used to pass through beam displacer (BD), polarisation state of $|V\rangle$ remains unchanged and polarisation state of $|H\rangle$ is shifted by 4mm. After $\mathrm{H_{6}}$ $(0^\circ)$, $\mathrm{H_{7}}$ $(45^\circ)$, $\mathrm{H_{8}}$ $(0^\circ)$ and $\mathrm{H_{9}}$ $(45^\circ)$, paths are divided into paths $a$ and $c$ with polarisation state $|H\rangle$, and paths $b$ and $d$ with polarisation state $|V\rangle$. Photons of the two polarization states are separated by PBS and photons are sent to the single photon detectors (SPDs) $\mathrm{D_{1}}$ and $\mathrm{D_{2}}$ through single-mode fibers.
Fiber couplers (FCs); long-wave path filter (LPF); dichroic half-wave plate (DHWP); dichroic polarizing beam splitter (DPBS); $\mathrm{H_{1}}$-$\mathrm{H_{12}}$, half-wave plates (HWPs); dichroic mirror (DM); $\mathrm{Q_{1}}$-$\mathrm{Q_3}$, quarter-wave plates (QWPs).}
\label{figure 3}
\end{figure*}

(ii) Delayed choice. In the experimental setup, the length of single-mode fiber for idler photons is extended by three meters compared to that for signal photons.
The measurement of idler photon (ancilla qubit-1) determines the initial state of signal photon (system qubit), and this measurement choice is made after the signal photon has been registered by the detector. The dual selections are made after signal photon has completely passed through the Sagnac interferometer and reached detector.
Therefore, the possibility that the photons already have knowledge of their paths before reaching the detector is eliminated.
This achievement aligns with the goal of the delayed-choice experiment.

(iii) Dual selections. In Figure~\ref{figure 3}, one can choose the measurement basis for idler photons from $\{|0\rangle_{a1},|1\rangle_{a1}\}$ and $\{|+\rangle_{a1},|-\rangle_{a1}\}$, which constitutes the selection of $\mathrm{BS_{in}}$. The measurement module for idler photons consists of $\mathrm{H_{3}}$ and a PBS, and by using different eigenbases for projective measurements, it determines the initial state of signal photons. After signal photons pass through quartz crystals, a relative phase $\varphi$ is added to their vertically polarized state $|1\rangle_{s}$. Upon passing through BS, the photons' paths splits into path A and path B, which introduces the ancilla-2 qubit $\{|0\rangle_{a2},|1\rangle_{a2}\}$. It should be noted that the system qubit is encoded by the polarization of signal photon, meanwhile the ancilla-2 qubit is encoded by the path of signal photon. Consequently, the BS utilized in our experiment does not correspond to either $\mathrm{BS_{in}}$ or $\mathrm{BS_{out}}$ as depicted in Figure~\ref{figure 1}. Here, we refer to Refs.~\cite{Tang1221,Rab1722}, which use an HWP with a rotation angle of $\mathrm{22.5^\circ}$ to achieve the effect of a Hadamard gate. The wave plates $\mathrm{H_{4}}$ ($22.5^\circ$) and $\mathrm{H_{5}}$ ($0^\circ$) placed in two paths realize the Hadamard gate and $\sigma_z$ gate on system qubit, respectively. Distinct paths indicate different quantum gates passed by photons, thereby implementing the interaction operator $\sigma_z\otimes|0\rangle_{a2}\langle0|_{a2}+H_3\otimes|1\rangle_{a2}\langle1|_{a2}$ between system qubit and ancilla-2 qubit. Then, we can select the absence or presence of $\mathrm{BS_{out}}$ according to the measurement outcomes of ancilla-2 qubit. Here, the interaction operator is slightly different from the controlled-$H$ gate in Figure~\ref{figure 2}, which replaces the identity operator with $\sigma_z$. However, it is not difficult to 
see that this alteration 
does not affect the experimental results since the measurement probabilities $P^{(i)}_{jk}$ remain the same.

Before delving into the discussion of experimental results, we first analyze the evolution of the quantum state of signal photons as they pass through experimental setups.
When idler photons are subjected to projective measurements with outcomes $|0\rangle_{a1}$, $|1\rangle_{a1}$, $|+\rangle_{a1}$, and $|-\rangle_{a1}$, the corresponding initial states of signal photons at the input port of Sagnac interferometer, depicted in Figure~\ref{figure 3}, are
$|\phi^{(0)}_{0}\rangle:=|0\rangle_{s}$, $|\phi^{(1)}_{0}\rangle:=\mathrm{e}^{i\varphi}|1\rangle_{s}$, $|\phi^{(+)}_{0}\rangle:=\frac{1}{\sqrt{2}}(|0\rangle_{s} + \mathrm{e}^{i\varphi}|1\rangle_{s})$ and $|\phi^{(-)}_{0}\rangle:=\frac{1}{\sqrt{2}}(|0\rangle_{s}- \mathrm{e}^{i\varphi}|1\rangle_{s}$), respectively.
 The reflected path (A) and the transmitted path (B) of photons are called $|1\rangle_{a2}$ and $|0\rangle_{a2}$, respectively. After passing through $\rm{H}_{4}$ and $\rm{H}_{5}$, the state evolves into
 \begin{eqnarray}
   |\psi^{i}_{sa2}\rangle = \frac{1}{\sqrt{2}} \left(iH_{3}|\phi^{(i)}_{0}\rangle|\mathrm{1}\rangle_{a2} + \sigma_{z}|\phi^{(i)}_{0}\rangle|\mathrm{0}\rangle_{a2} \right),
 \end{eqnarray}
where $H_{3}$ and $\sigma_z$ are the Hadamard gate and Pauli-z operator acting on the polarization states, respectively. $\{|\phi^{(i)}_{0}\rangle\}$ represents the polarization input states of Sagnac interferometer with $i=0,1,+,-$.
After two BDs and $\mathrm{H_{6}}-\mathrm{H_{9}}$, photons' paths are divided into four $a$, $b$, $c$ and $d$, they are called $|a\rangle_{a2}$, $|b\rangle_{a2}$, $|c\rangle_{a2}$ and $|d\rangle_{a2}$. The polarization state of single photons in paths $a$ and $c$ is $|0\rangle_{s}$, and the polarization state of single photons in paths $b$ and $d$ is $|1\rangle_{s}$. At this point, the signal photons state obtained at the output ports of the Sagnac interferometer in Figure~\ref{figure 3}, based on different projective measurements of idler photons, 
yields the corresponding probabilities at the detected ports:
\begin{eqnarray}
   && P^{(0)}_{a}=\dfrac{1}{4}, \ \ \  P^{(0)}_{d}=0,\\
   && P^{(1)}_{a}=\dfrac{1}{4}, \ \ \  P^{(1)}_{d}=\dfrac{1}{2},\\
   && P^{(+)}_{a}=\dfrac{1+\mathrm{cos}\varphi}{4}, \ \ \ P^{(+)}_{d}=\dfrac{1}{4},\\
   && P^{(-)}_{a}=\dfrac{1-\mathrm{cos}\varphi}{4}, \ \ \ P^{(-)}_{d}=\dfrac{1}{4}.
\end{eqnarray}
Here, $P^{(i)}_{a}$ and $P^{(i)}_{d}$ are the probabilities measured in path $a$ and $d$, respectively, and $i=0,1,+,-$ correspond to the project measurements on the idler photons with outcomes being  $|0\rangle_{a1}$, $|1\rangle_{a1}$, $|+\rangle_{a1}$ and $|-\rangle_{a1}$.

The number of photons detected in paths $a$, $b$, $c$ and $d$ represents the four projective measurement outcomes of the two-qubit state $|\psi^{\prime i}_{sa2}\rangle$. Specifically, we have $P^{(i)}_{a}=P^{(i)}_{01}$, $P^{(i)}_{c}=P^{(i)}_{11}$, $P^{(i)}_{b}=P^{(i)}_{00}$ and $P^{(i)}_{d}=P^{(i)}_{10}$. Therefore, the photons detected in paths $a$ and $c$ correspond to the experimental results when $\mathrm{BS_{out}}$ is present, while those detected in the other two paths correspond the results when $\mathrm{BS_{out}}$ is absent. Since paths $a$ and $c$ (see also $b$ and $d$) have the same physical significance, we choose to observe photons only on paths $a$ and $d$.
The SPDs are at port $a$ and $d$. Each SPDs only a fraction of the total state.
 Note that the signals from detectors $\mathrm{D_{1}}$ and $\mathrm{D_{2}}$ have to be delayed with approximately 11.5 ns, to perform coincidence counting with the signal of $\mathrm{D_{3}}$, as idler photons need longer time than signal photons to reach the detectors. The coincidence window is set to 1 ns. The presence or absence of $\mathrm{BS_{in}}$ and $\mathrm{BS_{out}}$ can be combined into four different configurations in Figure~\ref{figure 1}, each of which we observe, and then we obtain experimental results that are compared with theoretical results.

\begin{figure*}
\centering
\includegraphics[width=0.8\textwidth]{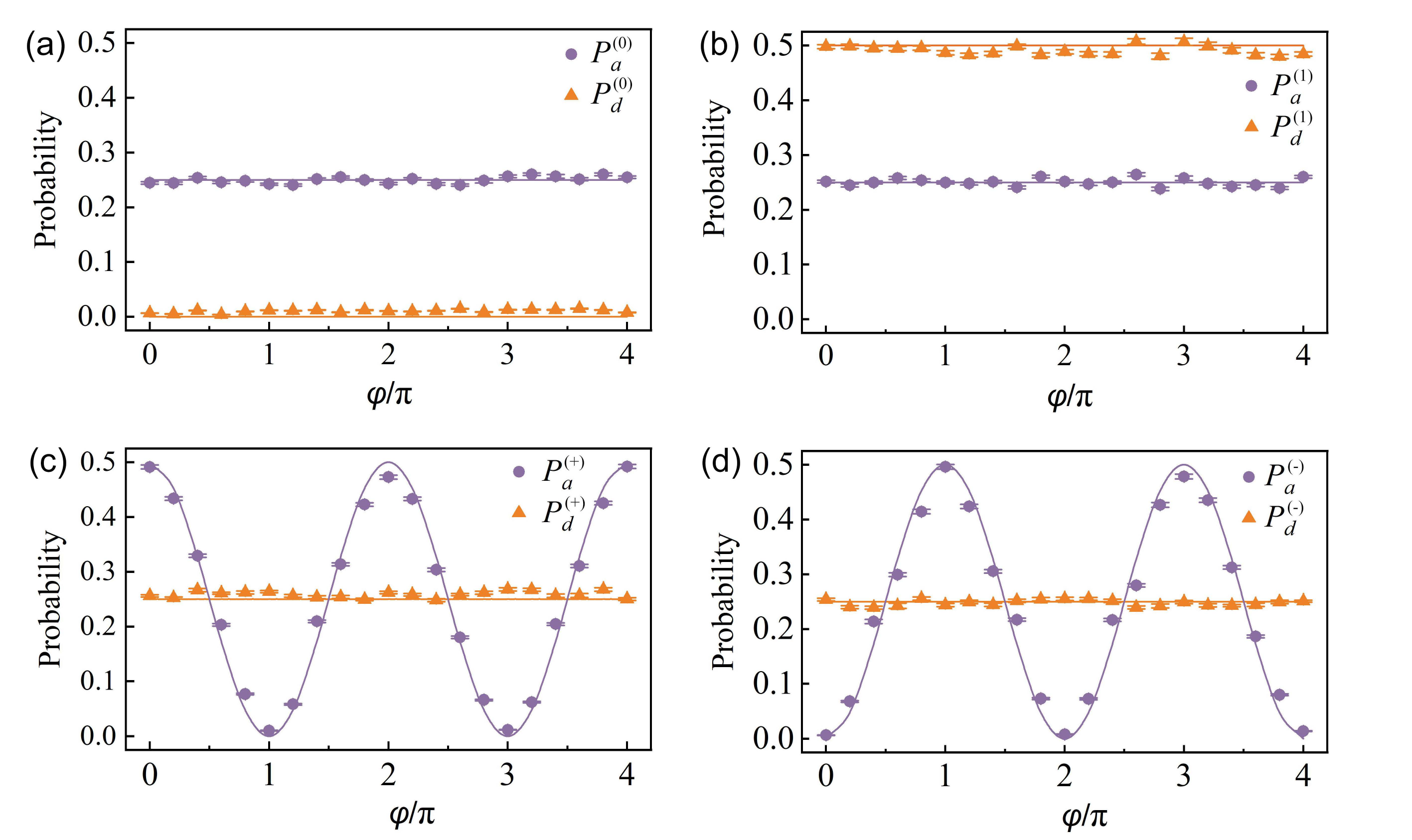}
\caption{The probabilities of measurement outcomes for varying phase $\varphi$. Purple dots represent experimental data for $P_{a}^{(i)}$, and purple solid lines are their theoretical values. Orange triangles represent experimental data for $P_{d}^{(i)}$, and orange solid lines are their theoretical values. Theoretically, $P_{a}^{(0)}=P_{a}^{(1)}=P_{d}^{(+)}=P_{d}^{(-)}=1/4$, $P_{d}^{(0)}=0$, $P_{d}^{(1)}=1/2$, and $P_{a}^{(\pm)}=(1\pm\mathrm{cos}\varphi)/4$. The error bars are derived from error propagation formula.
(a) $\mathrm{H_{3}}$ in Figure~\ref{figure 3} is set to $0^\circ$, and projective measurements are performed on idler photons with outcome states being $|0\rangle_{a1}$. (b) $\mathrm{H_{3}}$ is set to $45^\circ$, and projective measurements are performed on idler photons with outcome states being $|1\rangle_{a1}$. (c) $\mathrm{H_{3}}$ is set to $22.5^\circ$, and projective measurements are performed on idler photons with outcome states being $|+\rangle_{a1}$. The interference visibility $\mathcal{V}$ for path $a$  is  $97.35\%$. (d) $\mathrm{H_{3}}$ is set to $-22.5^\circ$, and projective measurements are performed on idler photons with outcome states being $|-\rangle_{a1}$. The interference visibility $\mathcal{V}$ for path $a$ is $95.87\%$.}
\label{figure 4}
\end{figure*}

\section{Experimental results}
The experimental probabilities $P^{(i)}_{jk}$ for different combinations of $\mathrm{BS_{in}}$ and $\mathrm{BS_{out}}$ selections are shown in Figure~\ref{figure 4}. In experiment, by varying parameter $\varphi$, we recorded the coincidence counts between detectors $\mathrm{D_{1}}$ and $\mathrm{D_{3}}$, as well as between $\mathrm{D_{2}}$ and $\mathrm{D_{3}}$, respectively.
 As discussed above, the selection of $\mathrm{BS_{in}}$ is controlled by the measurement basis of idler photon, while the paths of signal photon determine the selection of $\mathrm{BS_{out}}$. In Figure~\ref{figure 4}, the purple points $P^{(i)}_{a}$ are represent the experimental results when the $\mathrm{BS_{out}}$ is present, while the orange points $P^{(i)}_{d}$ are the experimental results when $\mathrm{BS_{out}}$ is absent. The absence of $\mathrm{BS_{in}}$ leads to particle-like behavior of photons, regardless of the presence or absence of $\mathrm{BS_{out}}$. This is evidenced by the lack of interference patterns in Figure~\ref{figure 4} (a) and (b). In contrast, Figure~\ref{figure 4} (c) and (d) depict scenarios where $\mathrm{BS_{in}}$ is present, and $\mathrm{BS_{out}}$ is inserted on path $a$ but not on path $d$.
 The simultaneous response of SPDs on two paths ($a$ and $d$). As depicted in Figure~\ref{figure 4} (c) and (d), path $a$ exhibits interference patterns, while path $d$ shows a lack of interference patterns.
 This suggests that, despite the presence of $\mathrm{BS_{in}}$, photons on path $d$ still exhibit particle-like behavior. Photons only exhibit wave-like properties when both $\mathrm{BS_{in}}$ and $\mathrm{BS_{out}}$ are present.

In our experiment, for each state of $\rm{BS_{in}}$, we are able to simultaneously observe the properties exhibited by photons under the two distinct conditions where $\rm{BS_{out}}$ is either present or absent. The experimental probabilities are determined by the ratio of the number of photons detected on individual paths ($N^{(i)}_{a}$ or $N^{(i)}_{d}$) to the total number of photons ($N_{t}$). Since we have $\sum_{i=0}^{1}(P^{(i)}_{a}+P^{(i)}_{d})=1$, the total number of photons can be obtained by summing the number of photons detected in paths $a$ and $d$ in both cases, i.e., $N_{t}=\sum_{i=0}^{1}(N^{(i)}_{a}+N^{(i)}_{d})$. For the case where the measurement basis of the idler photon is $\{|+\rangle_{a1},|-\rangle_{a1}\}$, the method for obtaining $N_{t}$ is similar.
Furthermore, the wave-like behavior can be quantified by the interference visibility $\mathcal{V}$, which is defined as $\mathcal{V}=(N_{\mathrm{max}}-N_{\mathrm{min}})/(N_{\mathrm{max}}+N_{\mathrm{min}})$,
where $N_{\mathrm{max}}$ and $N_{\mathrm{min}}$ are the maximum and minimum photon counts observed, respectively. Analysis of Figure~\ref{figure 4} (c) and (d) reveals that the highest interference visibility achieved in our experiment is 97.4\%.

If the rotation angles of $\rm{H}_3$-$\rm{H}_5$ are set within the range of $0^\circ$ to $22.5^\circ$, for paths $a$ and $c$, only $\rm{BS_{in}}$ is in a quantum superposition state; whereas for paths $b$ and $d$, both $\rm{BS_{in}}$ and $\rm{BS_{out}}$ are in quantum superposition states. To illustrate with a specific example, when the rotation angle of $\rm{H}_3$ is set to $15^\circ$, the quantum state of signal photons at the input port of Sagnac interferometer depicted in Figure~\ref{figure 3} is $\frac{\sqrt{3}}{2}|0\rangle_{s} + \frac{1}{2}\mathrm{e}^{i\varphi}|1\rangle_{s}$. When the rotation angles of $\rm{H}_4$ and $\rm{H}_5$ are set to $22.5^\circ$ and $15^\circ$ respectively, we can deduce that the measurement probabilities at the output ports $a-d$ are $P_{a}=(2+\sqrt{3}\cos \varphi)/8$, $P_{c}=(2-\sqrt{3}\cos \varphi)/8$, $P_{b}=(5+3\cos \varphi)/16$ and $P_{d}=(3-3\cos \varphi)/16$, respectively. The probabilities for all four paths exhibit interference patterns. However, compared to the $P_{a}^{(+/-)}$ interference pattern in Figure~\ref{figure 4}. This phenomenon indicates that the photons exhibit incomplete wave-like and particle-like behavior, meaning that photons are in a superposition state of wave-like and particle-like properties.

As shown in Figure~\ref{figure 3}, the optical path difference between two arms of Sagnac interferometer is very small. After the introduction of BDs, we ensure that when photons are in input state $|0\rangle$, their trajectories are precisely aligned with the path of Sagnac interferometer. Potential sources of experimental error may include the imperfect preparation of entangled state, the imperfect separation of horizontal and vertical polarization by BDs and PBS, and inaccuracies in the rotational adjustment of the half-wave plates $\mathrm{H_{6}}-\mathrm{H_{9}}$. As a result, photons traveling through paths A and B are not completely separated. Other errors between experimental and theoretical results are due to background counts, device instability, etc. The errors present in the experiment may cause the measured values of $P_{d}^{(0)}$ in Figure~\ref{figure 4} (a) to never be 0.
To reduce experimental errors, we perform $n = 230-500$ measurements to obtain each experimental dot, and
the error bars are calculated based on the error propagation formula.
Since the actual BS used does not achieve the ideal 50$:$50 
reflection-to-transmission ratio, to further reduce experimental errors, we choose to have both arms of Sagnac interferometer undergo one process of reflection and transmission.

\section{Discussions and conclusion}
This work builds upon and extends the insights from previous delayed-choice experiments. Additionally, based on the quantum delayed-choice experiment with dual selections, the implementation of a ``quantum eraser" experiment is possible. When idler photons are subjected to projective measurements onto $|+\rangle$ or $|-\rangle$, adjusting the rotation angle of $\rm{H_5}$ from $0^\circ$ to $22.5^\circ$ can transform $\rm{BS_{out}}$ from its initial absent state to a present state. This allows for the erasure of path information previously acquired by photons, enabling the observation of whether interference pattern can be restored. The ``quantum eraser" experiment has been extensively applied in the fields of quantum key distribution, two-photon imaging, and quantum sensing ~\cite{Schwindt99,Kim00,Bergou18,Weisz14,Chiou24,Kim23,Peng14,F10,B24,Ta}. Specifically, in Ref.~\cite{G17}, this is realized using entangled two-photon ghost imaging technology 
combined with Fourier transform methods.

In summary, we have realized a quantum Wheeler's delayed-choice experiment to observe wave-particle duality of single photons. In our experiments, we have used entangled photons, and implemented a delayed choice of measurement basis for idler photons, which controls the path of signal photons. This choice determines the absence or presence of the first beam splitter $\mathrm{BS_{in}}$ of the MZI. And the second choice has also been performed for the second beam splitter $\mathrm{BS_{out}}$ of the MZI. We can simultaneously observe different behaviors corresponding to the presence and absence of $\mathrm{BS_{out}}$, which greatly facilitates comparative analysis.

\section*{Acknowledgments}
This work is supported by the National Natural Science Foundation of China (Grants No.~62475127 and No.~11821404), Zhejiang Provincial Natural Science Foundation of China (Grant No. LZ25A040006), the Innovation Program for Quantum Science and Technology (Grant No. 2021ZD0301200).


\clearpage

\end{document}